\documentstyle[aps,preprint,tighten,epsfig,graphics,graphicx]{revtex}

\newcommand{\csch}{\,{\rm csch}\,}
\newcommand{\cG}{{\cal G}}

\begin{document}

\draft

\title{Casimir effect for the scalar field
under Robin boundary conditions: A functional integral approach}
\author{Luiz C de Albuquerque$^{1,}$\footnote{Email: \tt lclaudio@fatecsp.br } 
and R M Cavalcanti$^{2,}$\footnote{Email: \tt rmoritz@if.ufrj.br } }
\address{$^{1}$Faculdade de Tecnologia de S\~ao Paulo - CEETEPS - UNESP \\
Pra\c{c}a Fernando Prestes, 30, 01124-060 S\~ao Paulo, SP, Brazil \\
$^{2}$Instituto de F\'{\i}sica, Universidade Federal do Rio de Janeiro \\
Caixa Postal 68528, 21941-972 Rio de Janeiro, RJ, Brazil}
\date{June 1, 2004}

\maketitle

\begin{abstract}
In this work we show how to define the action of a scalar field in a such a way that 
Robin boundary condition is implemented dynamically, i.e., as a consequence of the 
stationary action principle. 
We discuss the quantization of that system via functional integration.
Using this formalism, we derive an expression for the Casimir
energy of a massless scalar field under Robin boundary conditions on
a pair of parallel plates, characterized by constants $c_1$ and $c_2$.
Some special cases are discussed; in particular, we show that
for some values of $c_1$ and $c_2$ the Casimir energy as a function
of the distance between the plates presents a minimum.
We also discuss the renormalization at one-loop order
of the two-point Green function in the $\lambda\phi^4$ theory
submitted to Robin boundary condition on a plate.
\end{abstract}


\section{Introduction}

The Casimir force between two uncharged macroscopic bodies in vacuum
is widely regarded as arising from the zero-point fluctuations
intrinsic to any quantum system. In the case of two 
flat parallel  plates a distance $a$ apart the pressure is proportional 
to $a^{-4}$ \cite{Casimir}.
In recent years several groups have performed high-precision measurements of
the Casimir force between a flat plate and a spherical 
surface (lens) or a sphere \cite{Exp1}, and also between two parallel 
flat plates \cite{Exp2}. In the last case the original Casimir formula
was confirmed to $15\%$ accuracy.

Due to its fundamental character the Casimir effect has applications
in many areas of physics, ranging from
the theory of elementary particles and interactions \cite{Bag,Extra}
to atomic and molecular physics \cite{long}. Besides, it has 
analogues in condensed matter physics, for instance in fluctuation
induced forces \cite{Kardar} and boundary critical phenomena \cite{Diehl}.
More recently, its relevance to the design and operation of micro- 
and nano-scale electromechanical devices has been emphasized \cite{Nano}.

Usually the details of the interaction between the vacuum fluctuations of the quantum  
field and the macroscopic bodies are neglected, and replaced by 
classical boundary conditions (BC) at the boundary of the latter. While Dirichlet
and Neumann BC have been extensively studied over the past, the more general
case of Robin BC has attracted little attention.

A field $\phi$ is said to obey Robin boundary condition at a
surface $\Sigma$ if its normal derivative at a point on $\Sigma$
is proportional to its value there:
\begin {equation}
\frac{ \partial }{ \partial n } \, \phi(x) = c\, \phi(x),
\qquad x \in \Sigma.
\end {equation} 
Neumann and Dirichlet boundary conditions are particular cases
of Robin boundary condition: the first one corresponds to $c = 0$;
the other is obtained in the limit $c \to \infty$ (assuming that
$\partial_n \phi$ is bounded).

The mixed case of Dirichlet-Robin (DR) BC were considered in 
\cite{Moste} for a 2D massless scalar field as a phenomenological 
model for a penetrable surface, with $c^{-1}$ playing the role of the 
finite penetration depth. Recently, the Casimir energy for a 
scalar field subject to Robin BC on one or two parallel planes
was computed in \cite{Romeo}.

Here we also compute the Casimir energy  
for a scalar field under Robin BC on two parallel planes.
However, we introduce a rather different approach which seems more amenable
to an eventual computation of radiative corrections in models containing 
interactions. Its starting point is the introduction of
suitable boundary terms in the action, which allows us to compute
the partition function of the system without the explicit 
imposition of Robin BC on the fields. In spite of that we show that
the two-point Green function does satisfy those boundary conditions.
We also find agreement with the main results of 
Ref.\ \cite{Romeo} for the Casimir energy, which was computed there using
the more conventional approach of summing the zero-point energy
of the normal modes of the field.
In particular, we show that for the mixed case of DR boundary conditions
the Casimir energy as a function of $a$ develops a minimum, i.e.,
there is a configuration of stable equilibrium.

Finally, we study the renormalization at one-loop order
of the two-point Green function in the $\lambda\phi^4$ theory
submitted to Robin BC on a plate. Our analysis differs
from previous ones \cite{Diehl,Symanzik} in two aspects:
(i) we keep $c$ arbitrary, instead of considering only the
the particular cases $c=0$ or $c=\infty$, and (ii)
we perform the regularization entirely in momentum space.
This procedure avoids dealing with distributions 
and test functions, which are unavoidable in the mixed
coordinate-momentum space regularization used in \cite{Diehl,Symanzik}. 

%%%%%%%%%%%%%%%%%%%%%%%%%%%%%%%%%%%%%%%%%%%%%%%%%%%%%%%%%%%%%%%%%%%%%%%%%%%%%%%%%

\section{The modified action}
\label{sec2}

Let us consider, for simplicity, a real scalar field living in
the half-space $z\ge 0$,\footnote{Conventions: $\hbar=c=1$, $x=({\bf x},z)$,
where ${\bf x}:=(x^0,\ldots,x^{d-1})$ and $z:=x^{d}$.}
satisfying the (Euclidean) equation of motion
\begin{equation}
\label{eqmotion}
-\partial^2\phi+U'(\phi)=0
\end{equation}
and subject to Robin boundary condition at $z=0$,
\begin{equation}
\label{RobinBC}
\partial_z\phi-c\phi\,\Big|_{z=0}=0.
\end{equation}
(We shall assume that $c\ge 0$, in order to avoid the possible appearance
of tachyons in the theory.)
One can easily verify that Eqs.\ (\ref{eqmotion}) and (\ref{RobinBC})
are consequences of the stationary action principle applied to the 
Euclidean action
\begin{equation}
\label{S1}
S[\phi]=\int d^dx\left\{\int_0^{\infty}dz\left[\frac{1}{2}\,(\partial_{\mu}\phi)^2
+U(\phi)\right]+\frac{1}{2}\,c\,\phi^2({\bf x},0)\right\},
\end{equation}
where
\begin{equation}
\int d^dx:=\lim_{\beta\to\infty}\lim_{L\to\infty}
\int_0^{\beta}dx^0\int_{-L/2}^{L/2}dx^1\cdots\int_{-L/2}^{L/2}dx^{d-1}.
\end{equation}
Indeed, computing $\delta S:=S[\phi+\eta]-S[\phi]$ up to second order in $\eta$
we obtain
\begin{equation}
\delta S=\int d^dx\left\{(-\partial_z\phi+c\phi)\,\eta\,\Big|_{z=0}
+\int_0^{\infty}dz\,[-\partial^2\phi+U'(\phi)]\,\eta\right\}+O(\eta^2),
\end{equation}
which implies (\ref{eqmotion}) and (\ref{RobinBC}) if $\phi$ is a 
stationary point of $S$.

Until now, we have been talking about a {\em classical} field.
What happens when one quantizes the theory? 
In the usual functional integrals approach one has to integrate over
all field configurations obeying certain boundary conditions.
If such is the case, is it necessary to retain the surface term
in the action? Bordag {\it et al}.\ \cite{Bordag}
argue that it is,
in order to ensure the Hermiticity (more precisely, the self-adjointness)
of the fluctuation operator
$\hat{\cal F}:=-\partial^2+U''(\phi_{\rm c})$, where $\phi_{\rm c}$ is the solution
to Eqs.\ (\ref{eqmotion}) and (\ref{RobinBC}). Saharian \cite{Saharian}
has also argued in favor of such a surface term: without it, the vacuum energy
evaluated as the sum of the zero-point energy of each normal mode of
the field does not agree with the result obtained by integrating
the vacuum energy density.

Here we propose a different approach. We retain the surface term
in the action, but we shall not impose any boundary condition at $z=0$
on the field configurations to be integrated over. 
Somewhat surprisingly, if we treat $U(\phi)$ as a perturbation,
the two-point Green function of the unperturbed theory does
satisfy Robin BC at $z=0$, i.e.,
\begin{equation}
\label{RobinBC2}
(\partial_z-c)\,\langle\phi(x)\,\phi(x')\rangle_0\,\Big|_{z=0}=0.
\end{equation}

Let us prove it. First of all, we write the partition function of
the unperturbed theory as\footnote{We follow here a procedure
very similar to the one employed in Ref.\ \cite{Aragao} in
the context of quantum field theory at finite temperature.}
\begin{equation}
Z_0=\oint[D\phi_1]\int_{\phi({\bf x},0)=\phi_1({\bf x})}[D\phi]\,\exp(-S_0),
\end{equation}
where $S_0$ is given by Eq.\ (\ref{S1}) without $U(\phi)$.
Note that we are integrating over all field configurations satisfying
the boundary condition $\phi({\bf x},0)=\phi_1({\bf x})$, and then
we integrate over all configurations of the surface field $\phi_1({\bf x})$.
In other words, we integrate over all possible boundary conditions at $z=0$.
$\phi({\bf x},z)$ also satisfies periodic BC in the {\bf x}-coordinates,
with period $\beta$ in the $x^0$-direction and $L$ in the others
(note, however, that we have the limits $\beta\to\infty$ and $L\to\infty$
in mind).

We now decompose $\phi$ as a sum of two fields: $\phi=\phi_0+\eta$,
where $\phi_0$ satisfies 
\begin{equation}
\label{phi0}
\partial^2\phi_0(x)=0,\qquad\phi_0({\bf x},0)=\phi_1({\bf x}),\qquad\phi_0({\bf x},\infty)=0.
\end{equation}
Note that, because of the boundary conditions imposed on $\phi_0$, $\eta$ vanishes at $z=0$
and $z=\infty$.  Eq.\ (\ref{phi0}) can be solved using Fourier 
transform; the result is
\begin{equation}
\label{Fphi0}
\phi_0({\bf x},z)=\int\frac{d^dk}{(2\pi)^d}\,e^{i{\bf k}\cdot{\bf x}}\,
\varphi_1({\bf k})\,e^{-kz},
\end{equation}
where $k=|{\bf k}|$ and $\varphi_1({\bf k})$ is the Fourier transform
of $\phi_1({\bf x})$.

In terms of the fields $\varphi_1$ and $\eta$, the partition
function $Z_0$ becomes the product of two independent functional
integrals: $Z_0=Z_A Z_B$, where
\begin{equation}
\label{Z1}
Z_A=\int[D\varphi_1]\,\exp\left\{-\int\frac{d^dk}{(2\pi)^d}
\,\frac{1}{2}\,(c+k)\,\varphi_1({\bf k})\,\varphi_1(-{\bf k})\right\},
\end{equation}
\begin{equation}
\label{Z2}
Z_B=\int_{\eta({\bf x},0)=0}[D\eta]\,\exp\left\{-\int d^dx
\int_0^{\infty}dz\,\frac{1}{2}\,(\partial_{\mu}\eta)^2\right\}.
\end{equation}

Let us now compute the two-point Green function. As a consequence
of the factorization of $Z_0$, one has $\langle\phi(x)\,\phi(x')\rangle_0=
\langle\phi_0(x)\,\phi_0(x')\rangle_A + \langle\eta(x)\,\eta(x')\rangle_B$.
According to Eq.\ (\ref{Z1}) we have
\begin{equation}
\langle\varphi_1({\bf k})\,\varphi_1({\bf k}')\rangle_A=
\frac{1}{c+k}\,\delta^{(d)}({\bf k}+{\bf k}').
\end{equation}
Combining this result with Eq.\ (\ref{Fphi0}) we obtain
\begin{equation}
\langle\phi_0(x)\,\phi_0(x')\rangle_A=
\int\frac{d^dk}{(2\pi)^d}\,e^{i{\bf k}\cdot({\bf x}-{\bf x}')}\,
\frac{e^{-k(z+z')}}{c+k}\,.
\end{equation}
On the other hand,
\begin{equation}
\langle\eta(x)\,\eta(x')\rangle_B=\int\frac{d^dk}{(2\pi)^d}\,
e^{i{\bf k}\cdot({\bf x}-{\bf x}')}\,D_{\eta}({\bf k};z,z'),
\end{equation}
where $D_{\eta}$ satisfies
\begin{equation}
\label{Deta}
(-\partial_z^2+k^2)\,D_{\eta}({\bf k};z,z')=\delta(z-z'),
\qquad D_{\eta}({\bf k};0,z')=D_{\eta}({\bf k};\infty,z')=0.
\end{equation}
One can easily verify that the solution to (\ref{Deta}) is given by
\begin{equation}
D_{\eta}({\bf k};z,z')=\frac{1}{k}\,\sinh(kz_{<})\,\exp(-kz_{>}),
\end{equation}
where $z_{<}(z_{>})={\rm min(max)}\{z,z'\}$. Collecting terms,
we finally obtain
\begin{equation}
\label{<>}
\langle\phi(x)\,\phi(x')\rangle_0=\int\frac{d^dk}{(2\pi)^d}\,
e^{i{\bf k}\cdot({\bf x}-{\bf x}')}\left[\frac{e^{-k(z+z')}}{c+k}
+\frac{1}{k}\,\sinh(kz_{<})\,\exp(-kz_{>})\right].
\end{equation}
One can easily verify that $\langle\phi(x)\,\phi(x')\rangle_0$ indeed
satisfies Robin BC at $z=0$, Eq.\ (\ref{RobinBC2}). 

%%%%%%%%%%%%%%%%%%%%%%%%%%%%%%%%%%%%%%%%%%%%%%%%%%%%%%%%%%%%%%%%%%%%%%%%%%%

\section{Casimir energy}

Let us now apply our procedure to the computation of the Casimir
energy of a free massless scalar field $\phi$ subject to Robin BC on two parallel
plates located at the planes $z=0$ and $z=a$,
\begin{equation}
\partial_z\phi-c_1\phi\,\Big|_{z=0}=0,\qquad
\partial_z\phi+c_2\phi\,\Big|_{z=a}=0\qquad(c_1,c_2\ge 0).
\end{equation}
The Euclidean action for such a system is given by
\begin{equation}\label{19}
S=\int d^dx\left\{\int_0^a  dz\left[\frac{1}{2}\,(\partial_{\mu}\phi)^2\right] +\frac{1}{2}\,c_1\phi^2({\bf x},0)+
\frac{1}{2}\,c_2\phi^2({\bf x},a)\right\},
\end{equation}
and its partition function is given by
\begin{equation}\label{20}
Z=\oint[D\phi_1][D\phi_2]\int_{\phi({\bf x},0)=\phi_1({\bf x})}^{\phi({\bf x},a)
=\phi_2({\bf x})}[D\phi]\,\exp(-S).
\end{equation}
The Casimir energy can be extracted from $Z$ using the identity
\begin{equation}
\label{lnZ}
E_0=-\lim_{\beta\to\infty}\frac{1}{\beta}\,\ln Z.
\end{equation}

As we did in the previous Section, we shall write $\phi$ as the
sum of two terms: $\phi=\phi_0+\eta$, where $\phi_0$ is the solution to
the classical equation of motion, $\partial^2\phi_0=0$, that obeys
the boundary conditions $\phi_0({\bf x},0)=\phi_1({\bf x})$ and 
$\phi_0({\bf x},a)=\phi_2({\bf x})$ (consequently $\eta({\bf x},0)=
\eta({\bf x},a)=0$).
By Fourier transforming in the ${\bf x}$-coordinates one can
explicitly solve for $\phi_0$, obtaining
\begin{equation}
\label{phi_0}
\phi_0({\bf x},z)=\int\frac{d^dk}{(2\pi)^d}\,\frac{e^{i{\bf k}\cdot{\bf x}}}
{\sinh ka}
\left[\varphi_1({\bf k})\sinh k(a-z)
+\varphi_2({\bf k})\sinh kz\right]
\end{equation}
where $ k=|{\bf k}|$ and $\varphi_j({\bf k})$ is the Fourier transform of
$\phi_j({\bf x})$, $j=1,2$.

Expressing $S[\phi]$ in terms of $\phi_0$ and $\eta$, we obtain 
$S=S_A+S_B$,
where
\begin{equation}
\label{SA}
S_A=\int d^dx\left\{\int_0^a dz\left[\frac{1}{2}\,(\partial_{\mu}\phi_0)^2\right]
+\frac{1}{2}\,c_1\phi_0^2({\bf x},0)+\frac{1}{2}\,c_2\phi_0^2({\bf x},a)\right\}
\end{equation}
\begin{equation}
S_B=\int d^dx\int_0^a dz\left[\frac{1}{2}\,(\partial_{\mu}\eta)^2\right]
\end{equation}
Since $\phi_0$ is a functional solely of $\phi_1$ and $\phi_2$,
the partition function $Z$ can be written as the product of two terms:
$Z=Z_A Z_B$, where
\begin{equation}
Z_A=\oint[D\phi_1][D\phi_2]\,\exp(-S_A),\qquad
Z_B=\int_{\eta({\bf x},0)=0}^{\eta({\bf x},a)=0}[D\eta]\,\exp(-S_B).
\end{equation}
It follows from Eq.\ (\ref{lnZ}) that the Casimir energy is
given by the sum of two terms, $E_0=E_A+E_B$,
of which the second one is the Casimir energy of a field 
subject to {\em Dirichlet} boundary conditions on the plates.
Since it is a well known result \cite{Ambjorn}, we shall just quote the result:
\begin{equation}
\label{EB}
E_B(a)=-\frac{L^{d-1}}{a^d}\,\Gamma\left(\frac{d+1}{2}\right)
(4\pi)^{-(d+1)/2}\,\zeta(d+1).
\end{equation}

Let us now compute $E_A$. Inserting (\ref{phi_0}) into
(\ref{SA}) we can rewrite $S_A$ in terms of $\varphi_1$
and $\varphi_2$ as
\begin{equation}
S_A=\frac{1}{2}\int\frac{d^dk}{(2\pi)^d}\,\varphi_i({\bf k})\,
M_{ij}({\bf k})\,\varphi_j(-{\bf k}),
\end{equation}
\begin{equation}
M_{ij}({\bf k})=(c_i+ k \coth ka)\,\delta_{ij}- k\csch ka\,(1-\delta_{ij}).
\end{equation}
Changing the variables of integration in $Z_A$ to
$\varphi_j$, we obtain
\begin{equation}
Z_A=\oint[D\varphi_1][D\varphi_2]\,\exp(-S_A[\varphi_1,\varphi_2])
=\prod_{\bf k} {\det}^{-1/2}M({\bf k}),
\end{equation}
hence
\begin{eqnarray}
\label{33}
E_A^{\rm bare}(a)&=&\lim_{\beta\to\infty}\frac{1}{2\beta}\sum_{\bf k}\ln\det M({\bf k})
\nonumber \\
&=&\frac{L^{d-1}}{2}\int\frac{d^dk}{(2\pi)^d}\,\ln\left[c_1c_2+ k^2
+(c_1+c_2) k\coth ka\right].
\end{eqnarray}
The expression above diverges, hence requires renormalization.
This is achieved by subtracting from it the quantity
\begin{equation}
\label{34}
\delta E_A=\lim_{a\to\infty}E_A^{\rm bare}(a)
=\frac{L^{d-1}}{2}\int\frac{d^dk}{(2\pi)^d}\,\ln[(c_1+k)(c_2+k)],
\end{equation}
which can be interpreted as part of the self-energy of the plates.
(An analogous subtraction is necessary in the calculation of $E_B$.)
Since $\delta E_A$ does not depend on the distance between the plates,
it does not contribute to the force between them. Its subtraction
from $E_A^{\rm bare}$ is thus permissible as long as one
is interested --- as we are --- only in the Casimir force.
The result of the subtraction is given by
\begin{equation}
E_{A}(a)=\frac{L^{d-1}}{2}\int\frac{d^dk}{(2\pi)^d}\,
\ln\left[1+\frac{2(c_1+c_2) k}{(c_1+ k)(c_2+ k)}
\cdot\frac{1}{e^{2ka}-1}\right].
\end{equation}
Performing the angular integration and adding the result to $E_B$,
Eq.\ (\ref{EB}), we finally obtain the Casimir energy for the massless
scalar field under Robin boundary conditions:
\begin{equation}
\label{ER}
E_0(c_1,c_2;a)=E_B(a)+\frac{L^{d-1}}{(4\pi)^{d/2}\,\Gamma\left(\frac{d}{2}\right)}
\int_0^{\infty}dk\,k^{d-1}\,\ln\left[1+\frac{2(c_1+c_2) k}{(c_1+ k)(c_2+ k)}
\cdot\frac{1}{e^{2ka}-1}\right].
\end{equation}

As a check of this result, we note that the integral vanishes
if $c_1=c_2\to\infty$ or $c_1=c_2=0$, thus reproducing the correct
result for Dirichlet-Dirichlet and Neumann-Neumann boundary conditions.
Dirichlet-Neumann boundary conditions ($c_1\to\infty$, $c_2=0$)
can also be treated exactly: in this case, the integral in Eq.\ (\ref{ER})
becomes
\begin{equation}
I(a):=\int_0^{\infty}dk\,k^{d-1}\,\ln\coth ka.
\end{equation}
Integrating by parts yields
\begin{equation}
I(a)=\frac{2a}{d}\int_0^{\infty}dk\,\frac{k^d}{\sinh 2ka}
=\frac{4a}{d}\sum_{n=0}^{\infty}\int_0^{\infty}dk\,k^d\,e^{-2(2n+1)ka}
=\left(2-\frac{1}{2^d}\right)\frac{\Gamma(d)}{(2a)^d}\,\zeta(d+1).
\end{equation}
Inserting this result into Eq.\,(\ref{ER}) and using the identity
$\Gamma(2z)=(4\pi)^{-1/2}\,2^{2z}\,\Gamma(z)\,\Gamma(z+1/2)$ 
\cite{Abramowitz} we finally obtain
\begin{equation}
E_0(\infty,0;a)=\frac{L^{d-1}}{a^d}\left(1-\frac{1}{2^d}\right)
\Gamma\left(\frac{d+1}{2}\right)(4\pi)^{-(d+1)/2}\,\zeta(d+1),
\end{equation}
which agrees with the correct result \cite{Krech}, thus giving us another check
of Eq.\,(\ref{ER}).

Next in simplicity are the following three cases:
(i) $c_1=c_2=c$, (ii) $c_1=\infty$, $c_2=c$, and (iii) $c_1=0$, $c_2=c$. 
We shall denote them RR, DR and NR, respectively (R = Robin, D = Dirichlet, 
and N = Neumann).
In all these cases, changing the variable of integration in Eq.\,(\ref{ER})
to $q=k/c$ allows us to rewrite it as
\begin{equation}
\label{Ealpha}
E_0^{\alpha}(c,a)=L^{d-1} c^d \,{\cal E}_{\alpha}(ca)\qquad(\alpha={\rm RR,DR,NR}),
\end{equation}
where
\begin{equation}
{\cal E}_{\alpha}(x)=-\frac{\Gamma\left(\frac{d+1}{2}\right)\,\zeta(d+1)}
{(4\pi)^{(d+1)/2}\,x^d}+\frac{1}{(4\pi)^{d/2}\,\Gamma\left(\frac{d}{2}\right)}
\int_0^{\infty}dq\,q^{d-1}\,\ln\left[1+\frac{f_{\alpha}(q)}{e^{2qx}-1}\right],
\end{equation}
with
\begin{equation} 
f_{\rm RR}(q)=\frac{4q}{(1+q)^2}\ ,\qquad f_{\rm DR}(q)=\frac{2q}{1+q}\ ,\qquad
f_{\rm NR}(q)=\frac{2}{1+q}\ .
\end{equation}
The graphs of ${\cal E}_{\alpha}(ca)$ in three spatial dimensions 
are depicted in Fig.\ \ref{fig1}. We can conclude
from it that the Casimir force between the plates: (i) is purely attractive
in the RR case (i.e., $c_1=c_2=c$); (ii) is repulsive at short distances and
attractive at long distances in the DR case, and (iii) is attractive at short 
distances and repulsive at long distances in the NR case. 

To understand
those behaviors, let us consider a free field subject to Robin BC at $z=0$,
i.e., $\partial_z\phi({\bf x},0)=c\phi({\bf x},0)$. If we write the $z$-dependent part of
$\phi$ as $\varphi(z)=\sin(kz+\delta)$, the previous equation becomes
$\tan\delta=k/c$. It follows that $\delta\to 0$ as $k\to 0$, and 
$\delta\to\pi/2$ as $k\to\infty$. In terms of $\varphi(z)$, this is 
equivalent to say that Robin BC tends to Dirichlet BC at low momentum,
and to Neumann BC at high momentum. In the jargon of renormalization
group theory, $c=\infty$ is an infrared and $c=0$ is
an ultraviolet attractive fixed point.

Let us now return to Fig.\ \ref{fig1}. According to the analysis above,
the RR curve should behave as the DD curve in the infrared 
(i.e., $a\to\infty$) and as the NN curve in the ultraviolet 
(i.e., $a\to 0$). In both cases,
the Casimir force is purely attractive, and so it is in the RR case.
The DR curve should behave as the DD curve as $a\to\infty$, and as the DN curve
as $a\to 0$; indeed, this is what we observe: attraction at long distances
and repulsion at short distances. The analysis of the NR curve is similar.

Such considerations suggest an interesting possibility. Let us suppose
that $0<c_1\ll c_2<\infty$. Then $E_0(c_1,c_2;a)\sim 
E_0(\infty,\infty;a)$ as $a\to\infty$, and 
$E_0(c_1,c_2;a)\sim E_0(0,0;a)$ as $a\to 0$. In both
these limits, therefore, the Casimir force is expected to be attractive.
However, since the crossover from a Dirichlet-like to a Neumann-like BC
takes place at different scales for each plate, there could be
a range of distances for which one has a Dirichlet-like BC at one plate 
and a Neumann-like BC at the other, thus leading to a repulsive force
between them. That such a possibility can indeed occur is shown in
Fig.\ \ref{fig2}.

%%%%%%%%%%%%%%%%%%%%%%%%%%%%%%%%%%%%%%%%%%%%%%%%%%%%%%%%%%%%%%%%%%%%%%%%%%

\section{Interacting field}

There is still another reason for the introduction of
boundary terms in the action: the perturbative
treatment of a renormalizable interacting theory requires boundary
counterterms in addition to the usual ones \cite{Diehl,Symanzik}. 
Thus, from a conceptual
point of view, it is more natural to treat boundary conditions
as resulting from the interaction of the field with a background
than to impose them {\it a priori}. We shall illustrate this point with
the calculation of the first order correction to the two-point Green function
of the $\lambda\phi^4$ theory in the presence of a flat boundary,
where the field is submitted to Robin BC.

The renormalized Euclidean Lagrangian density of the theory is given 
by\footnote{The factor of $\frac{1}{2}$ in the boundary term in the
action (\ref{S1}) is due to the fact that 
$\int_0^{\infty}\delta(z)\,f(z)\,dz=\frac{1}{2}\,f(0)$.}
\begin{equation}
{\cal L}=\frac{1}{2}\,(\partial_{\mu}\phi)^2+\frac{1}{2}\,m^2\phi^2
+c\,\delta(z)\,\phi^2+\frac{\lambda}{4!}\,\phi^4+{\cal L}_{\rm ct},
\end{equation}
with the field $\phi$ living in the half-space $z\ge 0$.
For simplicity, we shall assume $m=0$ and $c\ge 0$.
${\cal L}_{\rm ct}$ contains the renormalization counterterms ---
the usual ones,
\begin{equation}
{\cal L}_{\rm ct,\,bulk}=\frac{\delta Z}{2}\,(\partial_{\mu}\phi)^2
+\frac{\delta m^2}{2}\,\phi^2+\frac{\delta\lambda}{4!}\,\phi^4,
\end{equation}
plus boundary conterterms, which we shall exhibit later.

As we have seen in Sec.\ \ref{sec2}, the unperturbed two-point 
Green function (Feynman propagator) is given by [see Eq.\ (\ref{<>})]
\begin{equation}
G_0(x,x'):=\langle\phi(x)\,\phi(x')\rangle_0
=\int\frac{d^dk}{(2\pi)^d}\,e^{i{\bf k}\cdot({\bf x}-{\bf x}')}\,\cG_0(k;z,z'),
\end{equation}
where, in the massless theory,
\begin{equation}
\cG_0(k;z,z')=\frac{e^{-k(z+z')}}{c+k}+\frac{1}{k}\,\sinh(kz_<)\,\exp(-kz_>).
\end{equation}
If we neglect boundary counterterms,
the first order correction to $\cG_0$ is given by
\begin{equation}
\label{dG1}
\cG_1(k;z,z')=-\int_0^{\infty}dw\,\cG_0(k;z,w)\,\Sigma_1(w)\,\cG_0(k;w,z'),
\end{equation}
where the one-loop self-energy, including the mass counterterm, is given by
\begin{eqnarray}
\Sigma_1(w)&=&\frac{\lambda}{2}\int\frac{d^dq}{(2\pi)^d}\,\cG_0(q;w,w)+\delta m^2
\nonumber \\
&=&\frac{\lambda}{2}\int\frac{d^dq}{(2\pi)^d}\left(\frac{e^{-2qw}}{c+q}
+\frac{1-e^{-2qw}}{2q}\right)+\delta m^2.
\end{eqnarray}
(An ultraviolet cutoff $\Lambda$ is implicit in the integral above and 
wherever necessary below.)
To fix $\delta m^2$ we impose that $\Sigma_1(w)$ be finite for $w>0$ 
and vanishes for $w\to\infty$ (the latter condition is equivalent
to requiring that the physical mass is zero when one is infinitely
far from the plate).
These conditions imply
\begin{equation}
\delta m^2=-\frac{\lambda}{2}\int\frac{d^dq}{(2\pi)^d}\,\frac{1}{2q},
\end{equation}
hence
\begin{equation}
\Sigma_1(w)=\frac{\lambda}{2}\int\frac{d^dq}{(2\pi)^d}\left(\frac{1}{c+q}
-\frac{1}{2q}\right)e^{-2qw}.
\end{equation}

Although $\Sigma_1(w)$ is now a well defined function of $w$ for $w>0$, this is not
enough to ensure the finiteness of $\cG_1$. Indeed, $\Sigma_1(w)\sim w^{-(d-1)}$
for $w\to 0$, causing the divergence of the integral in Eq.\ (\ref{dG1}) for $d\ge 2$.
This problem can be (partially) solved by the boundary counterterm
\begin{equation}
\label{L2}
{\cal L}_{\rm ct,\,boundary}^{(1)}=\delta c\,\delta(z)\,\phi^2.
\end{equation}
It adds a new term to the self-energy
$\Sigma_1(w)$, turning it into 
$\widetilde{\Sigma}_1(w)=\Sigma_1(w)+2\delta c\,\delta(w)$. 
Equation (\ref{dG1}) then gets replaced by
\begin{equation}
\label{dG2}
\cG_1(k;z,z')=-\int_0^{\infty}dw\,\cG_0(k;z,w)\,\Sigma_1(w)\,\cG_0(k;w,z')
-\delta c\,\cG_0(k;z,0)\,\cG_0(k;0,z').
\end{equation}
Let us assume that $0<z<z'$; then Eq.\ (\ref{dG2}) yields
\begin{equation}
\label{dG3}
\cG_1(k;z,z')=-\delta c\,\frac{e^{-k(z+z')}}{(c+k)^2}
-\frac{\lambda}{2}\int\frac{d^dq}{(2\pi)^d}\left(\frac{1}{c+q}-\frac{1}{2q}\right)
[I(0,z)+I(z,z')+I(z',\infty)],
\end{equation}
where
\begin{equation}
I(a,b):=\int_a^b dw\,e^{-2qw}\,\cG_0(k;z,w)\,\cG_0(k;w,z').
\end{equation}
Since we are interested in the large-$q$ behaviour of the integrand,
we may restrict our attention to $I(0,z)$ ---
$I(z,z')$ and $I(z',\infty)$ are exponentially supressed in that limit. 
Performing the integration, we obtain
\begin{eqnarray}
I(0,z)&=&e^{-k(z+z')}\int_0^z dw\,e^{-2qw}\left[
\frac{e^{-kw}}{c+k}+\frac{1}{k}\,\sinh(kw)\right]^2
\nonumber \\
&\stackrel{q\to\infty}{\sim}&\frac{e^{-k(z+z')}}{(c+k)^2}\left[\frac{1}{2q}+\frac{c}{2q^2}+O\left(\frac{1}{q^3}\right)\right].
\end{eqnarray}
If we further multiply this result by the term in parentheses in Eq.\ (\ref{dG3}),
we conclude that the integrand in that equation behaves for large $q$ as
\begin{equation}
\frac{e^{-k(z+z')}}{(c+k)^2}\left[\frac{1}{4q^2}-\frac{c}{4q^3}+O\left(\frac{1}{q^4}\right)\right]
\stackrel{q\to\infty}{\sim}\frac{e^{-k(z+z')}}{(c+k)^2}\left[\frac{1}{4q(q+c)}+O\left(\frac{1}{q^4}\right)\right].
\end{equation}
Therefore, if we choose 
\begin{equation}
\label{dc}
\delta c=-\frac{\lambda}{8}\int\frac{d^dq}{(2\pi)^d}\,\frac{1}{q(q+c)}+\delta\overline{c},
\end{equation}
with $\delta\overline{c}$ finite, $\cG_1(k;z,z')$ becomes finite for $d<4$.
 
To see why the boundary conterterm (\ref{L2}) only solves the problem
partially, let us take $z=z'=0$ in Eq.\ (\ref{dG2}). 
A straightforward calculation then shows that
\begin{eqnarray}
\cG_1(k;0,0)&=&-\frac{1}{(c+k)^2}\left[\delta c+\frac{\lambda}{2}\int\frac{d^dq}{(2\pi)^d}
\left(\frac{1}{c+q}-\frac{1}{2q}\right)\frac{1}{2(q+k)}\right]
\nonumber \\
&=&-\frac{\delta\overline{c}}{(c+k)^2}+\frac{\lambda}{8(c+k)}\int\frac{d^dq}{(2\pi)^d}\,
\frac{1}{q(q+c)(q+k)}.
\label{dG4}
\end{eqnarray}
The result is finite for $d=2$, but diverges for $d=3$. In this case, another
boundary counterterm is needed. As we show below, it is given by
\begin{equation}
\label{L3}
{\cal L}_{\rm ct,\,boundary}^{(2)}=\delta b\,\delta(z)\,\phi(\partial_n-c)\phi,
\end{equation}
with $\partial_n$ the interior normal derivative (in our case, $\partial_n=\partial_z$).
Indeed, such a counterterm gives rise to an extra contribution to
$\cG_1(k;0,0)$, given by
\begin{eqnarray}
\Delta{\cG}_1(k;0,0)&=&-\delta b\int_0^{\infty}dw\,\delta(w)\,(\partial_w-2c)\,\cG_0^2(k;0,w)
\nonumber \\
&=&-\delta b\int_0^{\infty}dw\,\delta(w)\,(\partial_w-2c)\,\frac{e^{-2kw}}{(c+k)^2}
\nonumber \\
&=&\frac{\delta b}{c+k}.
\end{eqnarray}
If we choose 
\begin{equation}
\label{db}
\delta b=-\frac{\lambda}{8}\int\frac{d^dq}{(2\pi)^d}\frac{1}{q^2(q+c)}+\delta\overline{b},
\end{equation}
with $\delta\overline{b}$ finite, we cancel the divergence in 
$\cG_1(k;0,0)$ for $d=3$. Indeed,
\begin{eqnarray}
\cG_1(k;0,0)+\Delta\cG_1(k;0,0)&=&\frac{1}{c+k}\left[
\delta\overline{b}-\frac{\delta\overline{c}}{c+k}
-\frac{\lambda k}{8}\int\frac{d^dq}{(2\pi)^d}\,
\frac{1}{q^2(q+c)(q+k)}\right]
\nonumber \\
&=&\frac{1}{c+k}\left[
\delta\overline{b}-\frac{\delta\overline{c}}{c+k}
+\frac{\lambda}{16\pi^2}\,\frac{k}{c-k}\,\ln\left(\frac{k}{c}\right)\right]
\qquad(d=3).
\label{G1d3}
\end{eqnarray}

The boundary counterterm (\ref{L3}) is ineffective if $z,z'>0$; in this case,
$\Delta\cG_1(k;z,z')$ is identically zero. (As a consequence, $\cG_1(k;z,z')$
is insentive to the choice of $\delta\overline{b}$ if $z,z'>0$.) 
On the other hand, that
counterterm is necessary if $z=0$ and $z'>0$ (or vice-versa);
in this case, it is possible to show that the same choice (\ref{db}) for $\delta b$
also ensures the finiteness of $\cG_1(k;z,z')$.

In order to fix the finite part of the boundary counterterms,
$\delta\overline{b}$ and $\delta\overline{c}$, one has to impose
a pair of renormalization conditions. A natural choice,
for it is satisfied at tree level, is given by
\begin{equation}
\cG(\kappa;0,0)=(c+\kappa)^{-1},
\end{equation}
\begin{equation}
\frac{d}{dk}\,\cG(k;0,0)\bigg|_{k=\kappa}=-(c+\kappa)^{-2},
\end{equation}
where $\kappa$ is some arbitrary, but nonzero, mass scale. 
%With this choice, one gets
%
%\begin{eqnarray}\label{99}
%&&\delta b= -\lambda J_1(0,c)-\lambda\left[\,
%(c+2\kappa)\,J'_1(\kappa,c)+
%\kappa^{2} J'_2(\kappa,c)\,\right]\\
%&&\delta c=-\lambda J_0(0,c) -\lambda (c+\kappa)\left[\,
%\kappa J_2(\kappa,c)+(c+2\kappa) J'_1(\kappa,c)+
%\kappa^{2} J'_2(\kappa,c)\,
%\right]. \label{100}
%\end{eqnarray}
%
%where
%
%\begin{equation}\label{81}
%J_n(k,c):=\frac{1}{8}\int \frac{d^d q}{(2\pi)^d}\,
%\frac{1}{q^n(q+c)(q+k)},
%\end{equation}
%
%and $J'_n(\kappa,c)=\partial J_n(k,c)/\partial k$ at $k=\kappa$.
%Comparing with (59) and (55) we identify the finite 
%counterterms (for $d<4$) $\delta\overline{b}_1$ and 
%$\delta\overline{c}_1$ with the terms in square brackets
%on the RHS of (\ref{99}) and (\ref{100}), respectively.
%For completeness, we quote the results for the divergent
%part of the counterterms, which may be written as
%
%\begin{eqnarray}
%&& \delta b=-\frac{\lambda}{16\pi^2}\,\ln\frac{\Lambda}{\kappa}~,\nonumber\\
%&& \delta c=\frac{\lambda}{16\pi^2}\,\left[\,
%c\,\ln\frac{\Lambda}{\kappa}-\Lambda\,\right]~.
%\end{eqnarray}

We end this section with some remarks:

(A) Because of the boundary counterterms, the renormalized two-point Green function
{\em does not} satisfy Robin BC at $z=0$ --- except for $c\to\infty$ (Dirichlet BC);
in this case, the boundary condition is preserved at each order in perturbation
theory \cite{Symanzik}. 

(B) Inclusion of another plate at $z=a$ does not affect significantly the
overall picture. New ultraviolet divergences arise, but the theory is made finite
with the same type of counterterms used in the case of a single plate. 

%%%%%%%%%%%%%%%%%%%%%%%%%%%%%%%%%%%%%%%%%%%%%%%%%%%%%%%%%%%%%%%%%%

\section{Conclusions}

\begin{figure}[hbt]
\begin{center}
%\resizebox{!}{5cm}{\includegraphics{robinfig1.eps}}
\includegraphics*[scale=1, viewport=-10 -5 1000 360]{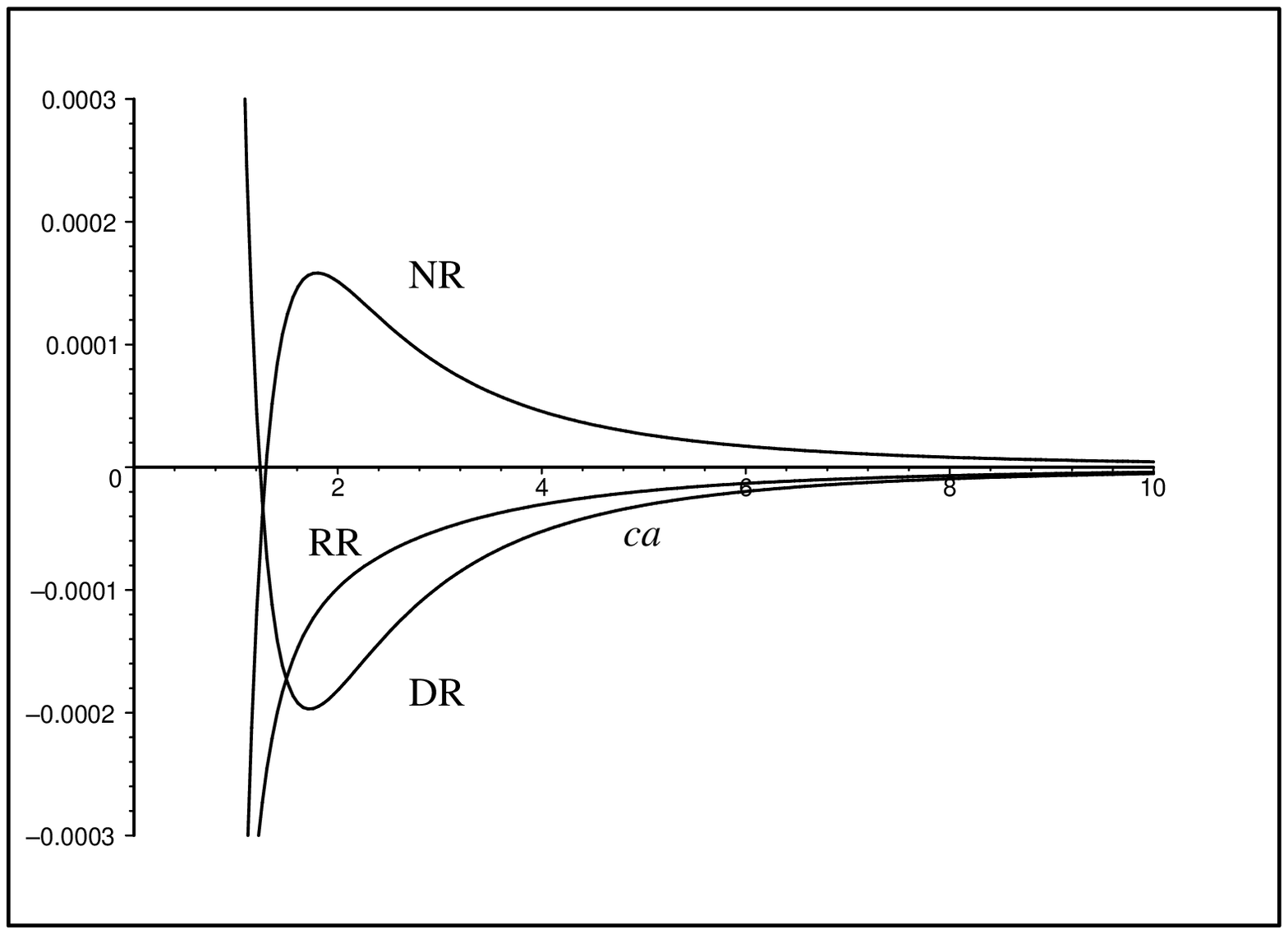}
\end{center}
\caption{${\cal E}_{\alpha}(ca)$ [see Eq.\ (\ref{Ealpha})] 
in three spatial dimensions for
$\alpha={\rm RR}$ ($c_1=c_2=c$), $\alpha={\rm DR}$ ($c_1=\infty$, $c_2=c$),
and $\alpha={\rm NR}$ ($c_1=0$, $c_2=c$).}%
\label{fig1}
\end{figure}

\begin{figure}[hbt]
\begin{center}
%\resizebox{!}{5cm}{\includegraphics{robinfig1.eps}}
\includegraphics*[scale=1, viewport=-10 -5 1000 360]{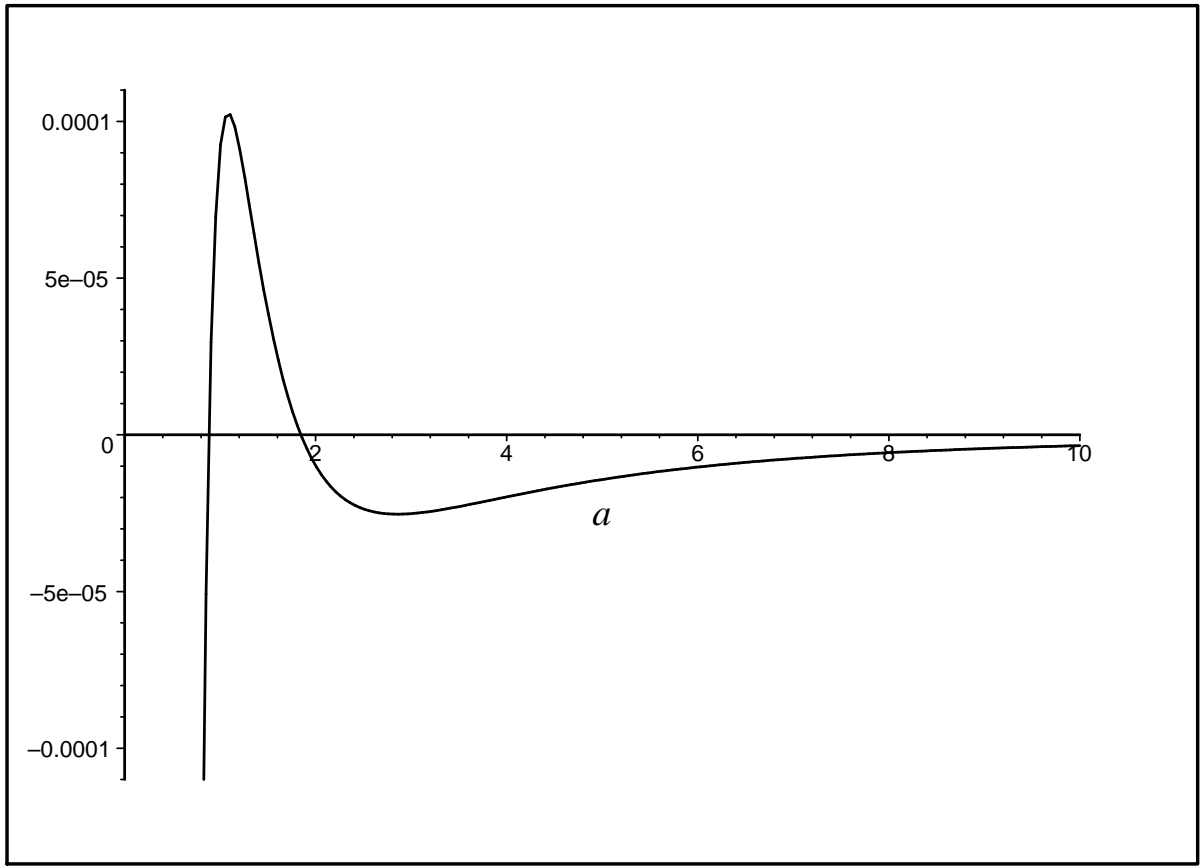}
\end{center}
\caption{Casimir energy per unit area {\it vs}.\ distance
between plates ($c_1=1/2$, $c_2=2$, $d=3$).}
\label{fig2}
\end{figure}

In this work we have computed the Casimir energy of a free massless scalar field
subject to independent Robin boundary conditions on
two parallel plates in $d$ spatial dimensions. 
It was shown that for mixed Dirichlet-Robin BC the 
Casimir energy as a function of the distance $a$ between the plates 
displays a minimum. 
We managed to understand the behavior of the Casimir energy as 
a function of $c$ relying on an analogy with the renormalization group flows
expected from infrared/ultraviolet fixed points. It was found that the 
Dirichlet BC ($c\to\infty$) is analogous to an attractive infrared fixed point 
whereas the Neumann BC ($c=0$) resembles an attractive ultraviolet fixed point. 
This interpretation is consistent with the numerical results shown 
in Fig.\ \ref{fig1} for the Casimir energy as a function of $ca$, 
and suggests that a crossover from 
Dirichlet-like to Neumann-like behaviors at the plates may lead to
a repulsive force between them.

We also provided a detailed analysis of renormalization 
for the two-point Green function $G(x,x')$ at 
first order in $\lambda$ in the $\lambda\phi^4$ theory. For simplicity
we worked out the case of Robin BC at a single flat boundary. 
We have shown that, in addition to the usual ``bulk'' counterterms,
one boundary counterterm is necessary to render $G$ finite
in the bulk, and a second one is necessary if at least one of 
its arguments lies on the boundary. This analysis is a necessary
step in the computation of radiative corrections to the Casimir
energy, which we intend to present elsewhere.

In view of the widespread interest on the Casimir effect and its
possible technological applications it is worthwhile to seek alternative 
computational tools which may go beyond or complement the existing
ones.
 % (for a recent proposal see \cite{Jaffe}).
For instance, in \cite{LCRIC} a resummation scheme to compute the
leading radiative corrections to the 
Casimir energy was suggested. On the other hand, the method outlined
in the present work seems well suited to the computation of radiative 
corrections either in a perturbative setting or eventually via
application of semi-classical methods.  Its starting point is the indirect 
implementation of the BC by means of appropriate terms in the action
functional, which is then reexpressed in terms of two kinds of fields: a 
field $\eta({\bf x},z)$ satisfying
Dirichlet BC at the surfaces, and two surface fields $\phi_j({\bf x})$ 
localized on the planes, depending only on the remaining $d$
transverse coordinates. The advantage of this procedure is that
the functional integration over the 
surface fields is unconstrained, i.e., one does not have to
enforce explicitly the Robin BC on the fields. 

The implementation of BC via local terms in the action is usually  
employed in  studies of boundary 
critical phenomena \cite{Diehl}. In that context, it can be shown that Dirichlet 
and Neumann BC correspond to the so-called ordinary ($c\to\infty$) and  
special transitions ($c=0$), respectively. The Robin BC is relevant
in the study of the crossover between those universality classes, 
for which, however, the computations become much more 
involved. It is relevant also for the analysis of
the ordinary transition; in this case, however, one may resort to an
expansion in powers of $c^{-1}$ \cite{Diehl}.
We expect that the methods proposed here can be useful to the study of 
the crossover for the relevant case of two flat planes.
This is presently under investigation.

%%%%%%%%%%%%%%%%%%%%%%%%%%%%%%%%%%%%%%%%%%%%%%%%%%%%%%%%%%%%%%%%%%%%%%%%%%%%

\acknowledgments
L.C.A. would like to thank the Mathematical Physics
Department of USP at S\~ao Paulo for its kind hospitality.
L.C.A.\ is partially supported by CNPq, grant 307843/2003-3.
R.M.C.\ is supported by CNPq and FAPERJ.

%\end{document}

%%%%%%%%%%%%%%%%%%%%%%%%%%%%%%%%%%%%%%%%%%%%%%%%%%%%%%%%%%%%%%%%%%%%

\end{document}